\documentclass{ws-ijmpd}
\usepackage[super,compress]{cite}
\begin{document}

\markboth{V.~Sahni and Y.~Shtanov} {Variable Gravity and the Faint Young Sun Paradox}

%
\catchline{}{}{}{}{}
%

\title{CAN A VARIABLE GRAVITATIONAL CONSTANT RESOLVE \\ THE
FAINT YOUNG SUN PARADOX\,?\footnote{This essay received an honorable mention in the 2014
Essay Competition of the Gravity Research Foundation.}  }

\author{VARUN SAHNI}

\address{Inter-University Centre for Astronomy and Astrophysics,\\
Post Bag 4, Ganeshkhind, Pune 411~007, India \\
varun@iucaa.ernet.in}

\author{YURI SHTANOV\footnote{Corresponding author.}}

\address{Bogolyubov Institute for Theoretical Physics,\\ 14-b Metrologicheskaya St.,
Kiev 03680, Ukraine \smallskip \\ Department of Physics, Taras Shevchenko National
University of Kiev,\\ 64/13 Vladimirskaya St., Kiev 01601, Ukraine \\
shtanov@bitp.kiev.ua}

\maketitle

\begin{history}
\received{Day Month Year} \revised{Day Month Year}
\end{history}

\begin{abstract}
Solar models suggest that four billion years ago the young Sun was $\sim 25\%$ fainter
than it is today, rendering Earth's oceans frozen and lifeless. However, there is ample
geophysical evidence that Earth had a liquid ocean teeming with life 4~Gyr ago. Since
${\cal L_\odot} \propto G^7M_\odot^5$, the Sun's luminosity ${\cal L_\odot}$ is
exceedingly sensitive to small changes in the gravitational constant $G$. We show that a
percent-level increase in $G$ in the past would have prevented Earth's oceans from
freezing, resolving the {\em faint young Sun\/} paradox. Such small changes in $G$ are
consistent with observational bounds on ${\Delta G}/G$. Since ${\cal L}_{\rm SNIa}
\propto G^{-3/2}$, an increase in $G$ leads to fainter supernovae, creating tension
between standard candle and standard ruler probes of dark energy. Precisely such a
tension has recently been reported by the Planck team.
\end{abstract}


\ccode{PACS numbers: 92.60.Iv, 92.70.Qr, 95.30.Sf, 96.60.Q-}


\section{The Faint Young Sun Paradox}

Between 4.5 and 3.9 billion years ago was an exciting time in the history of our planet.
This period witnessed the assembly of the Earth from planetesimals, the creation of the
moon, the formation of Earth's primeval oceans, etc. This was also when the Sun became a
main-sequence star.

The early Earth, however, faced a severe problem---that of the {\em faint young
Sun}.\cite{lunine,gough} Numerical models of the Sun's interior suggest that solar
luminosity about 4~Gyr ago (the Archean epoch), was only about $75\%$ of its present
value. This would have affected our oceans dramatically, reducing their temperature to
$268^\circ$K and ensuring that they were completely {\em frozen\/} during much of the
Archean.\cite{sagan72}

However, geophysical and climatological data do not support this view. On the contrary,
all evidence points to a temperate climate on Earth with liquid oceans teeming with
primordial life forms $\sim 4$~Gyr ago. Thus, the presence of metamorphosed sedimentary
rocks (from the Archean) shows strong evidence of erosion by liquid water. More evidence
comes from fossils of early life forms which depended upon water to survive.\cite{lunine}

Evidence for a temperate climate on Earth also comes from the abundance of stable
isotopes of Oxygen, $^{16}$O, $^{17}$O and $^{18}$O\@. Water on our planet is a mixture
of H$_2{}^{16}$O, H$_2{}^{17}$O and H$_2{}^{18}$O, with the former being the most
abundant. However, H$_2{}^{16}$O is also preferentially evaporated from our oceans which
should, consequently, have been enriched in $^{18}$O during the prolonged ice ages which
accompanied a fainter Sun. The $^{16}$O/$^{18}$O ratio, therefore, records Earth's
temperature through prehistory.

\begin{figure}[pb]
\centerline{\psfig{file=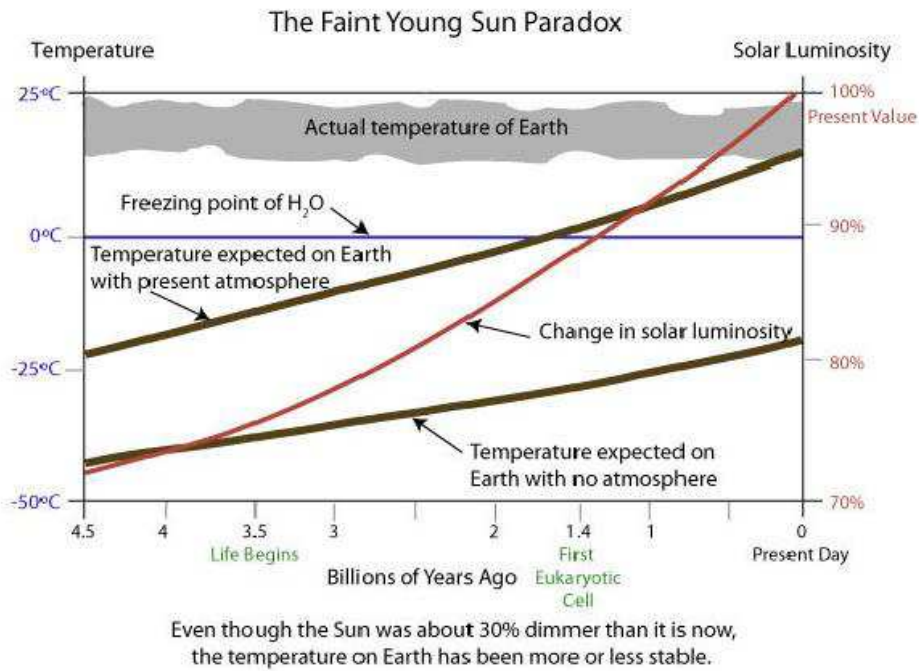,width=\textwidth}}
\vspace*{8pt}
\caption{Illustration of the Faint Young Sun Paradox. \label{fig:Reinforcement}}
\end{figure}

Luckily, relics of tiny sea creatures ``coccolithophorids'' can preserve the primordial
$^{16}$O/$^{18}$O ratio through their calcite (CaCO$_3$) shells, which absorb $^{16}$O
and $^{18}$O in equal amounts. Similarly, {\em cherts\/}, rocks composed of silica, also
bear testimony to the $^{16}$O/$^{18}$O ratio in the water in which these rocks
precipitated. Both the chert and CaCO$_3$ data imply that the Earth's temperature has
remained roughly unchanged over the past $4$~Gyr, which challenges the faint young Sun
hypothesis;\cite{knauth92,knauth05} see Fig.~\ref{fig:Reinforcement}.

\section{A Greenhouse Effect on Early Earth\,?}

While several arguments have been advanced to resolve the {\em Faint Young Sun
Paradox\/}, none is problem-free. For instance, one might try to compensate for the
diminished sunlight by assuming that Earth's early atmosphere had a far greater
concentration of greenhouse gases, such as CO$_2$, than at present.

However this hypothesis runs into difficulties. In order to maintain the Earth's
temperature above the freezing point of water during the early Archean, the abundance of
CO$_2$ in Earth's primeval atmosphere should have been {\em a thousand times\/} its
present value. One must, therefore, account for the enormous decline in the CO$_2$
concentration: from roughly $10$ atmospheres 4~Gyr ago, to $3 \times 10^{-3}$ atmospheres
at present.

While CO$_2$ can be removed from the atmosphere by bacteria and plants during
photosynthesis and also by the weathering of rock, the enormous concentrations referred
to earlier seem difficult to account for. Indeed, large amounts of CO$_2$ in the early
atmosphere would have led to the formation of the iron carbonate based compound {\em
siderite\/} (FeCO$_3$)\@. However, analysis of billion year old paleosols challenges this
picture by finding no FeCO$_3$. Found instead are iron silicates which support a much
more moderate presence of CO$_2$ in the early atmosphere. Moreover, it would be difficult
to make CO$_2$ disappear almost entirely from the Earth's atmosphere since much of it,
after reacting with rocks and being used by shell-forming organisms to form CaCO$_3$
shells, gets deposited on the ocean floor and makes its way back into the atmosphere via
volcanic activity and plate tectonics.\cite{lunine,Falkowski,KA} The presence of other
greenhouse gases is equally problematic.\cite{rosing,goldblatt}

\section{A Varying Gravitational Constant}

The seeming contradiction between a faint young Sun on the one hand, and support for a
temperate climate on early Earth (and Mars) on the other, can easily be reconciled if one
notes that the Sun's luminosity is exceedingly sensitive to the value of the
gravitational constant. Indeed, as first pointed out by Teller,\cite{teller,gamow} the
luminosity of the Sun, ${\cal L_\odot}$, depends upon its mass, $M_\odot$, and Newton's
gravitational constant, $G$, as ${\cal L_\odot} \propto G^7M_\odot^5$. Increasing the
value of $G$, one increases the solar luminosity. Moreover, a larger value of $G$ affects
the Earth's orbit by bringing it {\em closer to the Sun\/}. The (conserved) specific
angular momentum on a circular orbit of radius $r$ with velocity $v$ is $vr$. Since
$v^2r^2 = GM_\odot r$, it follows that the radius of the Earth's (almost circular) orbit
$r \propto 1/GM_\odot$. Since flux varies with distance as ${\cal F}_\odot \propto
r^{-2}$, the amount of sunlight received on Earth becomes
\begin{equation}
{\cal F}_\odot \propto G^9M_\odot^7, \label{eq:lum2}
\end{equation}
which translates into $T \propto {\cal F}_\odot^{1/4} \propto G^{2.25}M_\odot^{1.75}$ for
Earth's temperature\cite{teller,gamow,uzan} (also see Ref.~\refcite{Iorio}). From
(\ref{eq:lum2}), one gets
\begin{equation}
\frac{\Delta {\cal F}_\odot}{{\cal F}_\odot} = 9 \frac{\Delta G}{G} + 7\frac{\Delta
M_\odot}{M_\odot}\, . \label{eq:lum3}
\end{equation}
Substituting $\Delta {\cal F}_\odot/{\cal F}_\odot = 0.25$, one finds that a modest
increase of $2\%$ in the value of the gravitational `constant' over a lookback time of
4~Gyr would largely compensate for the diminished sunlight on Earth. This scenario, then,
would allow for Earth's temperature to remain roughly constant upto the present, in
agreement with current data and without invoking the need for injecting enormous amounts
of greenhouse gases into Earth's early atmosphere, for which no compelling evidence
exists.\cite{rosing}

Our estimate of $\Delta G / G \simeq 0.02$ at a lookback time of $\sim 4$~Gyr is in
excellent agreement with the current observational constraints on varying $G$, including
the results of Wu and Chen,\cite{wu-chen} who obtain  $-0.083 < \Delta G/ G < 0.095$ at
Recombination, and those of Accetta {\em et al.\@\/},\cite{accetta} who obtain $-0.3 <
\Delta G/ G < 0.4$ at Nucleosynthesis. Since the above estimate of $\Delta G / G$ is {\em
model independent\/}, one cannot directly relate it to the current value of ${\dot G}/G$.
Nevertheless, if one assumes, for simplicity, $G \propto t^{-\beta}$, then one finds
${\dot G}/{G}\big\vert_{t_0} \simeq - 4\times 10^{-12}\,{\rm yr}^{-1}$, which is in good
agreement with binary pulsar data,\cite{binary} millisecond pulsar data,\cite{milli} ages
of globular clusters\cite{globular} and lunar ranging experiments.\cite{lunar}

\section{Theories with Varying $G$ }

The idea that $G$ may be a function of time was first suggested by Dirac who postulated
$G \propto t^{-1}$ as part of his {\em Large Numbers Hypothesis\/}.\cite{dirac} Although
such a large variation in $G$ was shown to be inconsistent with observations, Dirac's
hypothesis was adopted into a field-theoretic framework in which Newton's constant
acquired the status of a {\em dynamical field\/}.\cite{jordan,brans}

The action for the gravity sector of such a {\em scalar-tensor\/} theory is\cite{uzan}
\begin{equation}\label{actionJF}
  S = \frac{1}{16\pi G_*}\int
     \Bigl[F(\varphi)R - Z(\varphi) (\partial \varphi)^2
        - 2U(\varphi)\Bigr] \sqrt{-g}\, {\rm d}^4 x \, ,
\end{equation}
where $G_*$ is the bare gravitational constant. Comparing this with the Einstein--Hilbert
action for general relativity,
\begin{equation}\label{GR}
 S = \frac{1}{16\pi G}\int R \sqrt{-g}\, {\rm d}^4 x \, ,
\end{equation}
one finds that the role of the gravitational constant is played by $G = G_*/F(\varphi)$,
where the scalar $\varphi$ satisfies the equation of motion
\begin{equation}
2Z(\varphi)~\Box\varphi = - \frac{dF}{d\varphi}\,R - \frac{dZ}{d\varphi}\,(\partial
\varphi )^2 + 2 \frac{dU}{d\varphi} \, . \label{eq:EOM}
\end{equation}
One therefore needs to solve (\ref{eq:EOM}) in order to determine the time-dependence of
$G\,[\phi(t)]$. This has indeed been done for specific choices of $U$ and $F$, including
$F = \xi\varphi^2$ (induced gravity) and $F = 1+\xi\varphi^2$ (non-minimal
coupling).\cite{uzan} Interestingly, for a suitable choice of $U$, such a model can
describe {\em dark energy\/} (DE) by making the universe accelerate at late
times.\cite{shinji_book,SH,PBM,Tsujikawa}

A varying gravitational constant also arises in higher-dimensional Kaluza--Klein
theories. For instance, the 5D Einstein--Hilbert action (described by the 5D  metric
$\bar g_{AB}$)
$$
 S=\frac{1}{12\pi^2 G_5}\int\bar R\sqrt{-\bar g}\, {\rm d}^5 x \, ,
$$
when compactified onto 4D via the decomposition
$$
 \bar g_{AB} = \left(
\begin{array}{cc}
  \displaystyle g_{\mu\nu}+\frac{A_\mu A_\nu}{M^2}\phi^2  & \quad \displaystyle \frac{A_\mu}{M}\phi^2 \medskip \\
  \displaystyle \frac{A_\nu}{M}\phi^2  & \phi^2 \\
\end{array}
 \right),
$$
results in the 4D action
\begin{equation}\label{KKaction}
 S=\frac{1}{16\pi G_*}\int\left(R - \frac{\phi^2}{4M^2}F^2\right)\phi\sqrt{-g}\, {\rm d}^4
 x\, .
\end{equation}
Comparing (\ref{KKaction}) with (\ref{GR}), one finds that $G \propto \phi^{-1}$, which
generalizes to $G \propto \phi^{-D}$ in the case of $D$ extra dimensions.

One should note that scalars such as $\phi$ are ubiquitous in string theory, where they
occur in the form of a dilaton whose vacuum expectation value determines the coupling
constants of string theory.\cite{witten} In such theories, which include SO(32) and
$E_8\times E_8$ heterotic theories, one usually expects the 4D coupling constants,
including $G$, to vary with time.

\section{Conclusions}

The {\em faint young Sun\/} hypothesis implies that during much of prehistory---from
4~Gyr to 1~Gyr ago---both Earth and Mars would have been frozen. While the presence of
water on early Earth is well documented, the presence of outflow channels and
phyllosillicates, discovered by the Mars Reconnaissance Orbiter and by the Mars
Exploration Rovers, also point to abundant liquid water on early
Mars.\cite{lunine,mckay91} Consequently, the faint young Sun presents a paradox for both
Earth and Mars.\cite{lunine,sagan72}

We have shown that tiny changes in $G$ can resolve the faint young Sun paradox by
increasing the Sun's luminosity and bringing the Earth closer to the Sun 4~Gyr ago.

A variable gravitational constant has other important consequences. The peak luminosity
of type Ia supernovae (SNIa) is proportional to the Chandrasekhar mass $M_{\rm Ch}$,
which depends upon Newton's constant. Consequently, ${\cal L} \propto M_{\rm Ch} \propto
G^{-3/2}$, and a larger value of $G$ in the past would result in SNIa being {\em
fainter\/} than predicted by the standard candle hypothesis \cite{Gaz,RU}. A varying $G$
would therefore imply {\em a new source of systematics\/} in SNIa data. This would show
up as tension between data sets probing DE using standard rulers (BAO\,+\,CMB) on the one
hand, and standard candles (SNIa) on the other, with the latter predicting a more {\em
phantom-like\,} equation of state for DE\@. Precisely such a tension has recently been
reported by the Planck team \cite{Planck}.

Models with variable $G$ may also resolve the {\em cosmological constant
problem\/}.\cite{weinberg} Self-tuning mechanisms can dynamically reduce the vacuum
energy to a small value. In doing so, they invariably turn the gravitational constant
into a {\em variable\/} quantity.\cite{Dolgov,Charmousis,Sola} Thus, the faint young Sun
paradox, the cosmological constant problem and dark energy may all be pieces of the same
puzzle requiring a common theoretical platform for their resolution.

\section*{Acknowledgments}

The authors acknowledge support from the India--Ukraine Bilateral Scientific Cooperation
programme. We acknowledge useful discussions with Adrian Melott and Prasad Subramanian.


\end{document}